\documentclass[%
 reprint,
 amsmath,amssymb,amsthm,
 nofootinbib,
 aps,
]{revtex4-1}

\usepackage{dcolumn}
\usepackage{bm}
\usepackage[mathlines]{lineno}
\usepackage{graphicx}
\usepackage{csquotes}

\begin{document}
\preprint{APS/123-QED}

\title{Consistent Descriptions of Quantum Measurement}

\author{Jianhao M. Yang}
\email{jianhao.yang@alumni.utoronto.ca}
\affiliation{Qualcomm, San Diego, CA 92121, USA}

\date{\today}

\begin{abstract}
The Wigner's friend type of thought experiments manifest the conceptual challenge on how different observers can have consistent descriptions of a quantum measurement event. In this paper, we analyze the extended version of Wigner's friend thought experiment (Frauchiger and Renner, Nature Comm. 3711, 9 (2018)) in detail and show that the reasoning process from each agent that leads to the no-go theorem is inconsistent. The inconsistency is with respect to the requirement that an agent should make use of updated information instead of outdated information. We then apply the relational formulation of quantum measurement to resolve the inconsistent descriptions from different agents. In relational formulation of quantum mechanics, a measurement is described relative to an observer. Synchronization of measurement result is a necessary requirement to achieve consistent descriptions of a quantum system from different observers. Thought experiments, including EPR, Wigner's Friend and it extended version, confirm the necessity of relational formulation of quantum measurement when applying quantum mechanics to composite system with entangled but space-like separated subsystems.  
\begin{description}
\item[PACS numbers] 03.65.Ta, 03.65.-w
\item[Keywords] Wigner's Friend Experiment, Quantum Measurement, Relational Quantum Mechanics
\end{description}
\end{abstract}
\maketitle

\section{Introduction}
\label{intro}
In quantum mechanics, the Wigner's friend~\cite{Wigner} thought experiment has been widely discussed as it tests the validity of many quantum interpretation theories. In this thought experiment setup, an observer (Wigner's friend) is placed inside a lab to perform a selective measurement of a quantum system using an apparatus system. She knows with certainty the measurement outcome based on the reading of a pointer variable of the apparatus. From her perspective, the quantum system has been projected into a definite state. Another observer, Wigner himself, is placed outside the lab. The entire lab, including Wigner's friend, is isolated from the rest of world. Hence Wigner describes the measurement process performed by his friend in the lab as a unitary time evolution. At the end of the experiment, from Wigner's perspective, the quantum system is in a superposition state that is entangled with the apparatus. Wigner does not know the measurement outcome. Thus, Wigner and his friend give two different descriptions of the same physical process happened inside the lab.

The interpretation of the situation created in the above thought experiment manifests the difference of various quantum theories. Wigner initially designed the thought experiment to argue that consciousness is a necessary component in the quantum measurement process. Deutsch further extended the thought experiment to be applicable to macroscopic system such as the lab system~\cite{Deutsch}. The intention of Deutsch's extension is to support the many-world interpretation of quantum mechanics~\cite{Everett, Wheeler57, DeWitt70}. According to the many-world interpretation, multiple branches of worlds are created when Wigner's friend performs the measurement. Each world has its own value of the measured variable. There is no wave function collapse. This is in contrast to the Copenhagen Interpretation (CI). According to CI, the superposition state of the measured system collapses into one of its eigenstate when measurement occurs. CI insists that the quantum description on the measurement process inside the lab depends on the measuring apparatus~\cite{Bohr, Bohr35, Jammer74}, thus the description is relative to the observer. Relational quantum mechanics (RQM)~\cite{Rovelli96, Rovelli07, Transs2018, Rovelli18, Yang2017, Yang2018} extends the spirit of CI and asserts that a quantum system must be described relative to another quantum system. RQM discards the separation of classical system and quantum system in CI and assumes all systems are quantum systems, including macroscopic systems. In RQM there is no absolute state for a quantum system, it is legitimate that Wigner and his friend have different accounts of the measurement process in the lab. Both RQM and Bayesian quantum mechanics (QBism)~\cite{Fuchs02, Fuchs13} consider wave function as a mathematical tool that encodes the observer's information of a quantum system. The so-called ``wave function collapse" is just an update of information based on actual measurement outcome. On the other hand, objective collapse theories suggest that the quantum state is objective and there is ontological element in the wave function. A superposed wave function will collapse randomly when the system reaches certain physical threshold~\cite{Penrose}. Thus, Wigner cannot assign a superposition state to the lab system at the end of the experiment~\cite{Wigner2}. However, the objective collapse theories imply that quantum mechanics is incomplete and require the Schr\"{o}digner Equation to be modified in some ways.

As we can see, the Wigner's friend thought experiment provides conceptual value to testify many quantum theories. Recently, Frauchiger and Renner proposed an extended version of Wigner's friend experiment (WFR experiment in short)~\cite{FR2018} to further manifests some of the conceptual difficulties. In the original Wigner's friend experiment, the different accounts between Wigner and his friend are not considered contradictory because they are based on different level of knowledge. It is always possible for Wigner to perform additional verification with his friend and find agreement on the measurement outcome. Thus, the two descriptions from Wigner and his friend are reconciled. The WFR experiment, however, creates a situation that at the end of some of the experiments (i.e., with a non-zero probability), such reconciliation is not possible. This imposes additional conceptual challenge for any quantum interpretation to address. In particular, Ref.~\cite{FR2018} proposes a no-go theorem, which states that three natural sounding assumptions cannot be all valid in the same time. The three assumptions are 1.) universal validity of quantum mechanics (Q), 2.) predictions from different observers are consistent (C), and 3.) a particular measurement only yields one single outcome, i.e., single world instead of many-world (S).

This paper gives a detailed analysis of the WFR experiment in the Schr\"{o}dinger picture by explicitly writing down the wave function each agent assigns to the composite system at different experiment step. The reason to use the Schr\"{o}dinger picture is that it is more convenient to analyze how the information encoded in the wave function is utilized in the reasoning process of each agent. One important rule in the reasoning process is that an agent should make use of the available information, no more and no less. The information can be that is encoded in a known wave function, or can be obtained through direct measurement result. However, in Section \ref{subsec:eWigner} we show that not every agent is reasoning by consistently following such rule.  

Nevertheless, there is significant conceptual value brought up by the extended Wigner's friend thought experiment as it provides a clear examples that in order to reconcile the different account between different agent, additional verification or communication is required. There is always possible to come up with another more complicate thought experiment to produce potential inconsistency. To completely resolve this issue, an operational principle for the reconciliation process is proposed in Section \ref{Resolution}. The principle is a necessary component to construct a description of a quantum system with complete available information. 

The Wigner's friend experiment and the extended version are yet another set of examples that manifest the conceptual values of the relational formulation of quantum measurement~\cite{Yang2018}, where quantum measurement is reformulated based on basic RQM principles~\cite{Rovelli96, Rovelli18} and one of RQM implementations~\cite{Yang2017}. Specifically, it is asserted~\cite{Yang2018} that for a given quantum system, description of its time evolution can be implicitly relative without calling out the observer, while description of a quantum operation must explicitly call out the observer. Information exchange is relative to a local observer in quantum mechanics. The assumption of a Super Observer\footnote{Super Observer refers to an observer who knows measurement results instantaneously from local observer from any location} should be abandoned, so as the notion of observer independent description of physical reality. It is shown~\cite{Yang2018} that the EPR paradox~\cite{EPR} can be resolved by abandoning the notion of observer independent description of physical reality. In addition, different local observers can achieve consistent descriptions of a quantum system if they are synchronized on the outcomes from any measurement performed on the system. The Wigner's friend paradox and its extended version confirm the necessity of synchronizing local measurement results. This is particularly true when an experiment involves multiple measurement steps. In conclusion, the EPR paradox and the Wigner's friend type of paradox serve as examples to confirm the conceptual value of the relational formulation of quantum measurement.

\section{The Extended Wigner's Friend Paradox}
\label{subsec:eWigner}
To make it easy for comparison and analysis, we will adopt the same notations used in Ref.~\cite{FR2018}. Fig.1 depicts a sketch of the WFR thought experiment. The time sequence is labeled by $t=n:ij$ where $n$ is the number of round in the experiment, $i$ labels the step within the round, and $j$ labels the sub-step. There are four agents in the WFR experiment. Agents $\overline{F}$ and $F$ are inside the lab $\bar{L}$ and $L$, respectively. Agents $\overline{W}$ and $W$ are outside the lab, and can perform measurement on $\bar{L}$ and $L$, respectively. System $R$ is a quantum randomness generator which outputs variable $r=tails$ or $r=heads$ with probability $2/3$ and $1/3$, respectively. System $S$ is a spin that is set to $|\downarrow\rangle$ if $r=heads$ and $|\rightarrow\rangle$ if $r=tails$. The experimental protocol is described in detailed in Box 1 of Ref.~\cite{FR2018}. Denote $D$ as the detector that agent $F$ uses to measure $S$, and $\bar{D}$ as the detector that agent $\overline{F}$ uses to measure $R$. Since the state of knowledge of an agent is always synchronized with the corresponding detector state, there is no need to distinguish them. We can simply define an apparatus composite system $A=D\otimes F$ and assign a quantum state to $A$. Similarly, $\bar{A}=\bar{D}\otimes \overline{F}$.

\begin{figure*}
\begin{center}
\includegraphics[scale=1.75]{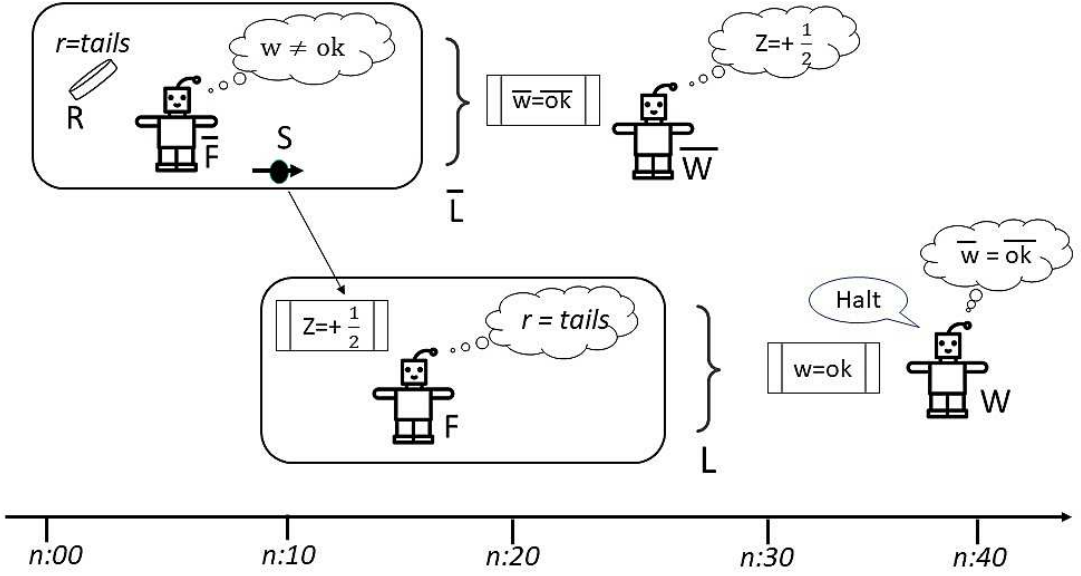}
\caption{Sketch of the extended Wigner's Friend thought experiment. More detailed description of the experiment protocol can be found in Figure 2 and Box 1 of Ref.~\cite{FR2018}. }
\label{fig:1}       
\end{center}
\end{figure*}

\subsection{Schr\"{o}dinger Picture}
In this subsection the WFR experiment is analyzed in Schr\"{o}dinger picture because it is more convenient to analyze how information is exchanged among the subsystems during the experiment. We first explicitly write down wave function for the composite system of the four sub-systems $R, S, A, \bar{A}$ in each step. Without loss of rigorousness, some of the wave functions in this section are not normalized. 

Before $t=n:00$, the wave function is initialized as
\begin{equation}
    \label{WF-1}
    \Psi^{init}_{all} = (\sqrt{\frac{1}{3}}|head\rangle_R + \sqrt{\frac{2}{3}}|tail\rangle_R)|init\rangle_{\bar{A},S,A}
\end{equation}
where $|init\rangle$ is an initial state for subsystems $\bar{A},S,A$. At step $n:00$, after time evolution, the wave function for the composite system becomes
\begin{equation}
    \label{WF0}
    \begin{split}
    \Psi^{00}_{all} &= U^{init\to 00}_{R\to \bar{L}S}\Psi^{init}_{all} =  (\sqrt{\frac{1}{3}}|head\rangle_R|\bar{h}\rangle_{\bar{A}}|\downarrow\rangle_S \\
    & + \sqrt{\frac{2}{3}}|tail\rangle_R|\bar{t}\rangle_{\bar{A}}|\rightarrow\rangle_S)|init\rangle_A
    \end{split}
\end{equation}
The subscript $all$ indicates all the four agents share the common knowledge of the initial state of the composite system. Between $t=n:00$ to $t=n:01$, agent $\overline{F}$ performs a projection measurement to $R$ and obtains result $|tail\rangle_R$. The resulting wave function (unnormalized) at $t=n:01$ is 
\begin{equation}
    \label{WF1}
    \begin{split}
    \Psi^{01}_{\bar{F}} & = |tail\rangle_R\langle tail|\Psi^{00}_{all}\rangle \\ 
    & = \sqrt{\frac{2}{3}}|tail\rangle_R|\bar{t}\rangle_{\bar{A}}|\rightarrow\rangle_S|init\rangle_A.
    \end{split}
\end{equation}
The probability of obtaining this results is given by
\begin{equation}
    \label{prob1}
    p(\bar{F}) = ||\langle tail|\Psi^{00}_{all}\rangle|| = \frac{2}{3}.
\end{equation}
Agent $\overline{F}$ then sends $S$ to $F$ in lab $L$, and $F$ performs measurement on $S$. During the measurement process, the composite system first goes through a unitary time evolution. At $t=1:0$, its wave function becomes
\begin{equation}
    \label{WF2}
    \begin{split}
    \Psi^{10}_{F\bar{F}} & = \sqrt{\frac{1}{2}}( |\uparrow\rangle_S|up\rangle_A+|\downarrow\rangle_S|down\rangle_A)|tail\rangle_R|\bar{t}\rangle_{\bar{A}}.
    \end{split}
\end{equation}
The subscript $F\overline{F}$ indicates that both agents $F$ and $\overline{F}$ share the same knowledge of this state information. After $F$ completes the measurement and obtains the result $z=1/2$, the wave function becomes
\begin{equation}
    \label{WF3}
    \begin{split}
    \Psi^{11}_{F} &= |\uparrow\rangle_S\langle \uparrow|\Psi^{10}_{F\bar{F}}\rangle \\
    & = \sqrt{\frac{1}{2}}|\uparrow\rangle_S|up\rangle_A|tail\rangle_R|\bar{t}\rangle_{\bar{A}}.
    \end{split}
\end{equation}
The subscript $F$ indicates only agent $F$ knows this state information. The probability of this measurement outcome is 
\begin{equation}
    \label{prob_F}
    p(F) = ||\langle\uparrow|\Psi^{10}_{F\bar{F}}\rangle|| = \frac{1}{2}.
\end{equation}

Now we consider the measurement outside the labs. When agent $\overline{W}$ performs the measurement, according to Ref.~\cite{FR2018}, the Heisenberg projector used is
\begin{equation}
    \label{pi}
    \pi^{n:00}_{(\overline{w},z)=(\overline{ok},-1/2)}=[(U^{init\to 00}_{R\to \bar{L}S})^\dag|\overline{ok}\rangle_{\bar{L}}|\downarrow\rangle_S][\cdot ]^\dag, 
\end{equation}
where $[\cdot ]^\dag$ denotes the adjoint of the operator defined in preceding factor, and
\begin{equation}
\label{pi}
   |\overline{ok}\rangle_{\bar{L}} = \sqrt{\frac{1}{2}}(|head\rangle_R|\bar{h}\rangle_{\bar{A}} - |tail\rangle_R|\bar{t}\rangle_{\bar{A}}).
\end{equation}
In Schr\"{o}dinger picture, this operator is equivalent to perform two projection operations on the wave function $\Psi^{00}_{all}$. First projector is $|\overline{ok}\rangle_{\bar{L}}\langle\overline{ok}|$, and the resulting wave function is
\begin{equation}
    \label{WF4}
    \begin{split}
        \Psi^{21}_{\overline{W}} &= |\overline{ok}\rangle_{\bar{L}}\langle\overline{ok}|\Psi^{00}_{all}\rangle \\
        & = |\overline{ok}\rangle_{\bar{L}}|\uparrow\rangle_S|init\rangle_A.
    \end{split}
\end{equation}
The probability for this measurement result is
\begin{equation}
    \label{p2}
    p(\overline{W}) = ||\langle\overline{ok}|\Psi^{00}_{all}\rangle || = \frac{1}{6}.
\end{equation}
The second projector is $|\downarrow\rangle_S\langle\downarrow|$ on $\Psi^{21}_{\overline{W}}$. Clearly the resulting wave function vanishes,
\begin{equation}
    \label{WF5}
    \Psi^{22}_{\overline{W}} = |\downarrow\rangle_S\langle\downarrow|\Psi^{21}_{\overline{W}}\rangle = 0.
\end{equation}
This enables agent $\overline{W}$ to confirm that $S$ must be in spin up state. 

Next we calculate the outcome of the measurement performed by agent $W$. According to Ref.~\cite{FR2018}, the Heisenberg projector is 
\begin{equation}
    \label{pi2}
    \pi^{n:00}_{(\overline{w},w)=(\overline{ok},ok)}=[(U^{init\to 00}_{R\to \bar{L}S})^\dag (U^{10\to 20}_{S\to L})^\dag|\overline{ok}\rangle_{\bar{L}}|ok\rangle_L][\cdot ]^\dag
\end{equation}
where
\begin{equation}
   |ok\rangle_L = \sqrt{\frac{1}{2}}( |\downarrow\rangle_S|down\rangle_A - |\uparrow\rangle_S|up\rangle_A).
\end{equation}
Since the unitary operator $U^{10\to 20}_{S\to L}$ and projector $|\overline{ok}\rangle_{\bar{L}}\langle\overline{ok}|$ commute, the overall projector in Eq.(\ref{pi2}) can be rearranged to
\begin{equation}
    \label{pi3}
    \pi^{n:00}_{(\overline{w},w)=(\overline{ok},ok)}=[(U^{init\to 00}_{R\to \bar{L}S})^\dag |\overline{ok}\rangle_{\bar{L}}(U^{10\to 20}_{S\to L})^\dag|ok\rangle_L][\cdot ]^\dag.
\end{equation}
Thus, in Schr\"{o}dinger picture, this operation is equivalent to apply operator $[(U^{10\to 20}_{S\to L})^\dag|ok\rangle_L][\cdot ]^\dag$ on the wave function $\Psi^{21}_{\overline{W}}$. Note that agent $\overline{W}$ announces the measurement results to agent $W$. Thus, both agents share the same knowledge of the wave function $\Psi^{21}_{\overline{W}}$. The unitary operator $U^{10\to 20}_{S\to L}$ evolves the wave function to
\begin{equation}
    \label{WF6}
    \Psi^{21}_{\overline{W}} \to \Psi^{30}_{W} = |\overline{ok}\rangle_{\bar{L}}|\uparrow\rangle_S|up\rangle_A.
\end{equation}
Then operator $|ok\rangle_L\langle ok|$ projects this wave function to
\begin{equation}
    \label{WF7}
    \begin{split}
        \Psi^{31}_W &= |ok\rangle_L\langle ok|\Psi^{30}_{W}\rangle \\
        & = \sqrt{\frac{1}{2}}|ok\rangle_L|\overline{ok}\rangle_{\bar{L}}.
    \end{split}
\end{equation}
The probability for this measurement result is
\begin{equation}
    \label{p3}
    p(W) = ||\langle ok|\Psi^{30}_{W}\rangle || = \frac{1}{2}.
\end{equation}
Therefore, the overall probability of the measurement represented by Eq.(\ref{pi2}) is
\begin{equation}
    \label{p3}
    p\{(\overline{w},w)=(\overline{ok},ok)\} = p(\overline{W})p(W) = \frac{1}{12}.
\end{equation}

\subsection{Reasoning of Each Agent}
With the wave functions for each step explicitly written down, we can examine how the reasoning of each agent works and how the reasoning leads to the no-go theorem in Ref.~\cite{FR2018}. Each agent can reason based on available knowledge on the wave function of the composite system, the predefined experiment protocol, and own measurement results.

Agent $\overline{F}$ is reasoning based on her knowledge of $\Psi^{10}_{F\bar{F}}$ after she completes the measurement on $R$ and obtained $|tail\rangle_R$. This wave function can be written as
\begin{equation}
    \label{WF8}
    \Psi^{10}_{F\bar{F}}  = |fail\rangle_L|tail\rangle_R|\bar{t}\rangle_{\bar{A}}
\end{equation}
where $|fail\rangle_L = \sqrt{\frac{1}{2}}( |\uparrow\rangle_S|up\rangle_A+|\downarrow\rangle_S|down\rangle_A)$,
which is orthogonal to $|ok\rangle_L$. Hence agent $\overline{F}$ predicts $W$ will observe $(w=fail)$ at time $t=n:31$. Agent $F$ is reasoning based on her measurement result on $S$ and knowledge of $\Psi^{11}_F$. Since $S$ is in the $|\uparrow\rangle$ state, agent $F$ infers that $\overline{F}$ obtains $(R=tail)$ according to the experiment protocol. Thus, $F$ is certain that $\overline{F}$ predicts that $W$ will observe $(w=fail)$ at time $t=n:31$. Applying assumption (C), agent $F$ is also certain that $W$ will observe $(w=fail)$ at time $t=n:31$.

Agent $\overline{W}$ is reasoning based on wave function $\Psi^{00}_{all}$ and his measurement result on the lab $\bar{L}$. Since the measurement result is $(\overline{w}=\overline{ok})$ and the resulting $\Psi^{20}_{\overline{W}}$ is orthogonal to $|\downarrow\rangle_S$, he infers that $S$ is in the $|\uparrow\rangle$ state. This further implies $F$ predicts with certainty that $W$ will observe $(w=fail)$ at time $t=n:31$. Again, by virtue of assumption (C), $\overline{W}$ is certain that $W$ will observe $(w=fail)$ at time $t=n:31$. Since $\overline{W}$ announces the measurement result $(\overline{w}=\overline{ok})$ to $W$, the same reasoning for $\overline{W}$ is applicable to $W$. Therefore, at the end of the reasoning, $W$ is certain that he will observe $(w=fail)$ at time $t=n:31$. 

But according to Eq.(\ref{p3}), the probability that $W$ will observe $(w=ok)$ and $\overline{W}$ will observe $(\overline{w}=\overline{ok})$ at time $t=n:31$ is $1/12$. This contradicts the end result of the reasoning described earlier. The no-go theorem is proved based on this contradiction. 

\subsection{Inconsistency in the Reasoning}
\label{critiques}
Let's take a deeper examination of the assumptions in the reasoning process, particularly how the available information is used. There are two naturally sounding rules here: 
\begin{enumerate}
    \item An agent should only make use of information available at hand. For instance, an agent should not use information from the measurement results of other agents unless there is explicit communication between the two agents.
    \item An agent should make use of all legitimate information available at hand. For instance, an agent should make use of latest information once her own measurement result is available, instead of still using updated information.
\end{enumerate}
We will examine whether each agent consistently follows these rules in their reasoning process.

\begin{table*}[ht]
\caption{The observations and reasoning of agents $F$ and $\overline{F}$. (This is part of Table 3 of Ref.~\cite{FR2018} which tabulates the reasoning of all agents.)}
\label{tab:1}       
\renewcommand*{\arraystretch}{1.4}
\begin{tabular}{|m{1cm}|m{1.7cm}|m{4.5cm}|m{4.5cm}|m{4.5cm}|}
\hline
 Agent & Observation & Statement inferred via (Q) & Further implied statement & Statement inferred via (C) \\
\hline
$\overline{F}$ & $r=tail$ at time $n:01$ & Statement $\overline{F}^{n:02}$: ``I am certain that $W$ will observe $w=fail$ at time $n:31$" & & .\\
\hline
$F$ & $z=+\frac{1}{2}$ at time $n:11$ &  Statement $F^{n:12}$: ``I am certain that $\overline{F}$ knows that $r=tail$ at time $n:01$". &  Statement $F^{n:13}$: ``I am certain that $\overline{F}$ is certain that $W$ will observe $w=tail$ at time $n:31$ & Statement $F^{n:14}$: ``I am certain that $W$ will observe $w=tail$ at time $n:31$\\
\hline
\end{tabular}
\end{table*}

First, agent $\overline{F}$ is reasoning based on her knowledge of wave function $\Psi^{10}_{F\bar{F}}$. She draws the conclusion that $W$ will observe ($w=fail$) at time $t=n:30$ by assuming if a projection measurement $|ok\rangle_L\langle ok|$ is performed on $\Psi^{10}_{F\bar{F}}$. The wave function $\Psi^{10}_{F\bar{F}}$ is changed at $t=n:21$ after agent $F$ performs the projection measurement $|\uparrow\rangle_S\langle\uparrow|$. However, agent $\overline{F}$ is an observer outside lab $L$, she does not know the measurement outcome performed by $F$ even though from the experiment protocol she knows that $F$ will perform a measurement before $t=n:30$. In other word, the information available to $F$ is not available to $\overline{F}$. Thus, it is legitimate for $\overline{F}$ to assume that information encoded in $\Psi^{10}_{F\bar{F}}$ stays the same~\footnote{There is time evolution from $t=n:10$ to $t=n:30$. However, time evolution does not change the correlation information encoded in the wave function. See Ref.~\cite{Yang2018}.} at time $t=n:30$. This assumption leads to the conclusion that statement $\overline{F}^{n:02}$ in Table 1 is true not only at $t=n:10$, but also at $t=n:31$. We conclude that the reasoning of agent $\overline{F}$ complies with the two rules mentioned earlier. 

Now consider the reasoning from agent $F$. After agent $F$ obtains the measurement result that $S$ is in the $|\uparrow\rangle$ state at $t=n:11$, agent $F$ infers that $\bar{F}$ obtains $(r=tail)$ based on her knowledge of $\Psi^{00}_{all}$. Thus, according to the reasoning in Table 1, agent $F$ concludes that $W$ will observe ($w=fail$) at time $t=n:31$. Note such conclusion depends on the reasoning of agent $\overline{F}$, who in turn depends on the information encoded in wave function $\Psi^{10}_{\bar{F}F}$. However, at $t=n:11$, $F$ knows precisely that the updated wave function is given by $\Psi^{11}_{F}$. If the operator $|ok\rangle_L\langle ok|$ is applied on $\Psi^{11}_{F}$, agent $F$ can conclude that $W$ will observe ($w=ok$) at time $t=n:31$ with probability of $1/2$, and the resulting wave function is
\begin{equation}
    \Psi^{12}_F = |ok\rangle_L\langle ok|\Psi^{11}_F\rangle  = \sqrt{\frac{1}{2}} |ok\rangle_L|tail\rangle_R|\bar{t}\rangle_{\bar{A}}.
\end{equation}
This is inconsistent with her own previous reasoning outcome. The reasoning of agent $F$ presented in Ref.~\cite{FR2018} does not follow Rule 2. She is implicitly based on the information encoded in wave function $\Psi^{10}_{\bar{F}F}$ instead of available updated information encoded in $\Psi^{11}_{F}$ after she performs the measurement on $S$. But if $F$ is reasoning based on $\Psi^{11}_{F}$, statement $F^{n:14}$ in Table 1 is not valid, and the proof of the no-go theorem becomes questionable.

Agent $\overline{W}$ is reasoning based on information encoded in wave function $\Psi^{00}_{all}$. If we consider the two labs $L$ and $\bar{L}$ as a whole, the action that agent $\overline{F}$ sends the physical copy of spin $S$ to $F$ is an internal interaction between the two labs. What happens inside the labs $L$ and $\bar{L}$ are unknown to $\overline{W}$. The measurement process carried by agent $F$ and $\bar{F}$ are described by agent $\overline{W}$ as time evolution such that system $R$ and $\bar{A}$ are entangled due to the measurement by $\bar{F}$, and subsystem $S$ and $A$ due to the measurement by $F$. Hence, at time $t=n:20$, from $\overline{W}$ point of view, the wave function should be
\begin{equation}
    \label{WF25}
    \begin{split}
    \Psi^{20}_{\overline{W}} & = U_{S\to L}^{10\to 20}U_{R\to \bar{L}_S}^{init\to 10}\Psi_{all}^{init} \\
    & = \sqrt{\frac{1}{3}}\{(|head\rangle_R|\bar{h}\rangle_{\bar{A}} + |tail\rangle_R|\bar{t}\rangle_{\bar{A}}) \\
    & \otimes |\downarrow\rangle_S|down\rangle_A 
     + |tail\rangle_R|\bar{t}\rangle_{\bar{A}}|\uparrow\rangle_S|up\rangle_A\}.
    \end{split}
\end{equation}
Agent $\overline{W}$ performs measurement at time $t=n:20$ with projector $|\overline{ok}\rangle\langle \overline{ok}|$ on wave function $\Psi^{20}_{\overline{W}}$, resulting in wave function
\begin{equation}
    \label{WF26}
    \Psi^{21}_{\overline{W}}=|\overline{ok}\rangle\langle \overline{ok}|\Psi^{20}_{\overline{W}}\rangle = |\overline{ok}\rangle|\uparrow\rangle_S|up\rangle_A.
\end{equation}
This leads to the same conclusion as Eq.(\ref{WF5}). Since $\overline{W}$ does not know the measurement results from $\bar{F}$ and $F$, the information encoded in $\Psi^{21}_{\overline{W}}$ is incomplete but legitimate according to the two reasoning rules. Similar statement can be applied to the reasoning of agent $W$.

In summary, the reasoning processes from agent $\overline{F}, {\overline{W}}$, and $W$ strictly follow the two rules mentioned earlier. However, the reasoning of agent $F$ violates Rule 2. On one hand, $F$ knows the measurement results on the spin system $S$. Thus, she knows the complete information on the system $S$ and updated wave function available to her is $\Psi^{11}_F$. On the other hand, the reasoning of $F$ presented in Ref.~\cite{FR2018} still relies on earlier information encoded in wave function $\Psi^{10}_{\bar{F}F}$. This inconsistency puts the proof of the no-go theorem in Ref.~\cite{FR2018} in question.

\section{RQM Resolution}
\label{Resolution}
\subsection{Synchronization of Measurement Result}
\label{sync}
Given the outcome of a measurement performed by a local observer $O_I$ on a quantum system $S$ is not necessarily available to another observer $O_E$, observer $O_E$ may assign $S$ a wave function that does not encode the complete information on $S$. Consequently $O_I$ and $O_E$ can have different descriptions of $S$. In the context of relational quantum mechanics~\cite{Rovelli96, Rovelli07, Transs2018, Rovelli18, Yang2017, Yang2018}, it is legitimate that different observers give different descriptions of a same quantum system because their level of knowledge on the system could be different. We will briefly describe RQM and its implication on quantum measurement in order to provide sufficient context for later discussion. 

In RQM, a quantum system is described relative to another reference system~\cite{Rovelli96}. The relational properties between two systems are more basic than the independent properties of a system. We recently proposed an implementation of RQM such that quantum mechanics can be reformulated with relational properties as starting point~\cite{Yang2017}. Ref.~\cite{Yang2018} applies such implementation to quantum measurement and further clarifies that while time evolution of a given quantum system can be described without explicitly calling out the observing system, a quantum measurement must be described explicitly relatively to the observing system. Quantum measurement is essentially a process to extract information from a quantum system using another measuring system\footnote{Strictly speaking, here the meaning of information refers to the correlation between the measured system and the measuring system.}. Such process should be described relative to the local observer. An observer who does not access to the measurement results will not have the complete information and can only describe the system up to the level of previous knowledge that the observer has. To ensure the descriptions of different observers are consistent, Ref.~\cite{Yang2018} proposes that different observers should synchronize information regarding the measurement results. This can be summarized as the following principle.
\begin{displayquote}
\textit{A complete description of a quantum system relative to an observer is achieved by taking into account of any quantum operation occurred to the system. To ensure consistent descriptions of a quantum system, measurement outcome obtained by a local observer must be communicated to other observers. }
\end{displayquote}
This principle appears quite intuitive. However, there are several subtleties that need further clarifications.
\begin{enumerate}
\item A quantum system may experience a long history of quantum measurements by different apparatuses over time. The principle does not say an observer needs to know the measurement outcome of every occurrence. Instead, suppose an observer knows the initial state of a system $S$ at time $t_0$, denoted as $\Psi_S(t_0)$, and the observer wants to give a complete description of $S$ at time $t_1$, denoted as $\Psi_S(t_1)$. The principle requires that the observer must know the outcome of any measurement on $S$ occurred between $t_0$ and $t_1$.
\item A quantum system can be a composite system that consists multiple subsystems. A measurement may be only applied to one of the subsystems. However, if the subsystems are entangled, measurement of any subsystem is considered as measurement of the entire system. For example, if a composite system has two entangled subsystems $A$ and $B$ that are remotely separated. Supposed observer $O_I$ near $A$ performed measurement on $A$. The result must be communicated to another observer $O_E$ near $B$ so that both observers have consistent descriptions of the composite system.
\item The principle is essentially an operational one. The synchronization mechanism between observers can be achieved through direct additional measurement on the system, as described in Ref.~\cite{Rovelli96}, or through some forms of information exchange between two observers. Such information exchange is achieved through physical interaction. The details of such process are not the main focus here but we assume it
 follows quantum mechanics principles.
\item When an observer receives the measurement outcome, he should update the wave function according to the measurement theory~\cite{Yang2018, Neumann, Nelson00}. Suppose the initial state of a system $S$ is $\Psi_0$, the measurement is described by an operator $\hat{M}_m$, and $\hat{M}_m$ is invariant when switching observers, then the wave function is updated to be
\begin{equation}
    \label{measureWF1}
    |\Psi_m\rangle = \frac{\hat{M}_m|\Psi_0\rangle}{\sqrt{\langle\Psi_0|\hat{M}_m^\dag \hat{M}_m|\Psi_0\rangle}}
\end{equation}

\end{enumerate}
Equivalently, this principle can be stated in the Heisenberg representation as following. Suppose the state of a quantum system $S$ is $|\Psi_S\rangle$, a complete Heisenberg operation $\pi(t_0 \to t_1)$ to describe the quantum events happened to $S$ between $t_0$ and $t_1$ must capture all intermediate operations that extract information from the system during this period. Note that an operation may be performed by a different observer. Missing an intermediate operation in the Heisenberg operator will result in incomplete description of $S$. 

The synchronization principle is not stated in the original RQM~\cite{Rovelli96}. However, in Section ~\ref{oRQM}, we will provide analysis that it is not conflicting with the basic RQM principles.

\subsection{The Resolution}
With the synchronization principle, we can proceed the reasoning in the extended Wigner's friend experiment to see if it leads to a contradiction.

To implement the synchronization principle, we need to modify the experiment protocol. For each measurement performed by an agent at $t=n:k0$ and completed at $t=n:k1$, we require the agent to communicate the measurement result to other agents. Suppose the communication is completed at $t=n:k2$ and all agents update their wave function of composite system accordingly. They should assign a same wave function to the composite system. In other words, at $t=n:k2$, there is only one wave function that is shared among all agents. 

At $t=n:00$, the wave function is given by $\Psi^{00}_{all}$ in Eq.(\ref{WF0}). At $t=n:01$, $\overline{F}$ completes her measurement and obtains outcome of $R=tail$ with probability of $2/3$. $\bar{F}$ sends the $S$ to $F$ physically and informs all other agent the measurement outcome. At $t=n:02$ all agents update the wave functions to $\Psi^{02} = \Psi^{01}$ given in Eq.(\ref{WF1}). The wave function goes through time evolution to $\Psi^{10}$ given in Eq.(\ref{WF2}). We omit the subscript for $\Psi$ since it is expected to be the same to all agents. 

At $t=n:11$, agent $F$ completes her measurement and obtains outcome of $S=1/2$ with probability of $1/2$, the resulting wave function is $\Psi^{11}$ given by Eq.(\ref{WF3}). Agent $F$ informs the outcome to other agents, and all agents update the wave functions at $t=n:12$ to $\Psi^{12}=\Psi^{11}$.

At $t=n:20$, agent $\overline{W}$ performs measurement and describes the process based on his knowledge of wave function $\Psi^{12}$. The resulting wave function with the measurement outcome of $\overline{w} = \overline{ok}$ is
\begin{equation}
    \label{WF20}
    \begin{split}
    \Psi^{22} & = |\overline{ok}\rangle_{\bar{L}}\langle\overline{ok}|\Psi^{12}\rangle \\
    & = \sqrt{\frac{1}{2}}|\uparrow\rangle_S|up\rangle_A|\overline{ok}\rangle_{\bar{L}},
    \end{split}
\end{equation}
with probability of $1/2$. $\overline{W}$ then communicates the result to agent $W$ and agent $W$ update the wave function to $\Psi^{22}$ at time $t=n:22$.

At $t=n:30$, agent $W$ performs measurement and describes the process based on his knowledge of wave function $\Psi^{22}$. The resulting wave function with the measurement outcome of $w = ok$ is
\begin{equation}
    \label{WF30}
    \begin{split}
    \Psi^{31} & = |ok\rangle_L\langle ok|\Psi^{22}\rangle \\
    & = \sqrt{\frac{1}{2}}|ok\rangle_L|\overline{ok}\rangle_{\bar{L}},
    \end{split}
\end{equation}
with probability of $1/2$. The overall probability to obtain the measurement outcome of $(\overline{w}, w) = (\overline{ok}, ok)$ from the initial wave function $\Psi^{00}_{all}$ is the product of the probabilities for the four measurement outcomes,
\begin{equation}
    \label{finalProb}
    p = \frac{2}{3}\times\frac{1}{2}\times\frac{1}{2}\times\frac{1}{2} = \frac{1}{12}.
\end{equation}
There is no contradiction or ambiguity in this reasoning process. 

The resolution can be explained in the Heisenberg representation as well. From agent $\overline{F}$ point of view, the complete Heisenberg projector used for reasoning to reach the statement of $(w=ok)$ is
\begin{equation}
    \label{pi5}
    \pi^{n:00}_{(w=ok)}=[(U^{init\to 00}_{R\to \bar{L}S})^\dag |tail\rangle_R (U^{10\to 20}_{S\to L})^\dag|ok\rangle_L][\cdot ]^\dag.
\end{equation}
Similarly, from agent $F$ point of view, the complete Heisenberg projector used for reasoning to reach the statement of $(w=ok)$ should be
\begin{equation}
    \label{pi6}
    \pi^{n:00}_{(w=ok)}=[(U^{init\to 00}_{R\to \bar{L}S})^\dag |tail\rangle_R (U^{10\to 20}_{S\to L})^\dag |\uparrow\rangle_S |ok\rangle_L][\cdot ]^\dag.
\end{equation}
If agent $F$ applies the operator in Eq.(\ref{pi6}) to the initial state, she will obtain a statement that agent $W$ will observe $(w=ok)$ with non-zero probability. Thus, agent $F$ cannot reach a conclusion that $W$ will observe $(w=fail)$ with certainty. Ref.~\cite{FR2018} just use $|\uparrow\rangle_S\langle\uparrow|$ to describe agent $F$'s reasoning process, but this operator is incomplete. 

\subsection{Entanglement}
The synchronization principle can be further understood with the concept of quantum entanglement. Entanglement measures the quantum correlation among different system~\cite{Horodecki}. When the subsystems of the composite system are entangled, each subsystem encodes information about other subsystems. Measurement on any of these subsystems extracts information of other subsystems. Thus, the measurement outcome of one subsystem should be communicated to observers who are local to other subsystems in order for them to have a complete description on their local subsystems. 

In the WFR experiment, the subsystems include $R, \bar{A}, S$ and $A$. The initial wave function $\Psi^{00}_{all}$ in Eq.(\ref{WF0}) shows that subsystems $R, \bar{A}$, and $S$ are entangled. The fact that agent $\overline{W}$ is able to deduce that $S$ is in the $|\uparrow\rangle$ state after he obtains measurement outcome of $\overline{w}=\overline{ok}$, is due to the entanglement information encoded in $\Psi^{00}_{all}$. As a consequence, the measurement outcome on subsystem $R$ should be communicated to agent $F$ who is interested in the quantum state of $S$, and to agent $W$ who is interested in the quantum state of lab $L$ that consists of $S$ and $A$. Certainly the result should be also communicated to agent $\overline{W}$ who is interested in the quantum state of lab $\bar{L}$ that consists of both $R$ and $\bar{A}$. 

Similarly, the wave function $\Psi^{10}_{\bar{F}F}$ at time $t=n:10$ shows that $S$ and $A$ are entangled. Measurement outcome obtained by agent $F$ should be communicated to agent $W$. On the other hand,  subsystems $S$ and $A$ are unentangled (i.e., in a product state) with subsystems $R$ and $\bar{A}$, as shown in $\Psi^{10}$, the measurement performed by $F$ will not give additional information of $R$ and $\bar{A}$. Since agent $\overline{W}$ is interested in measuring $\bar{L}$ which consists both $R$ and $\bar{A}$, it is not absolutely necessary for agent $\overline{W}$ to obtain the measurement outcome from agent $F$. He will predict the same measurement result of lab $\bar{L}$ with or without the information. This can be seen by applying projector $|\overline{ok}\rangle_{\bar{L}}\langle ok|$ on either wave function $\Psi^{10}$ or $\Psi^{11}$. The reason we demand agent $F$ to communicate the result to agent $\overline{W}$ is that the final goal of the experiment is to obtain measurement outcome of both ($\overline{w}=\overline{ok}$) and ($w=ok$). Such measurement goal needs complete description of all the four subsystems.

\section{Discussion and Conclusion}

\subsection{Operational and Conceptual Implications}

The relational formulation of quantum measurement results in two implications of the quantum measurement.
\begin{enumerate}
    \item \textit{Measured reality is relative.} Information obtained through quantum measurement is local. Measurement must be described explicitly relative to the local observer.  
    \item \textit{Synchronization of local reality.} This is essentially the synchronization principle discussed in Section \ref{sync}.
\end{enumerate}
In traditional quantum mechanics, these two implications of quantum measurement are not applicable due to the assumption of Super Observer. Quantum mechanics was initially developed as a physical theory to explain results of observation of microscopic systems, for instance, spectrum of light emitted from hydrogen atoms. In such condition, the observed system as a whole is much smaller than the apparatus. An observer can practically read the results from different subsystems at the same time. The assumption of Super Observer becomes operational even though it is conceptually incorrect. However, when one wishes to construct a quantum theory for composite system with entangled subsystems that are spatially separated by distance that is larger than the typical measuring apparatus by orders of magnitude, the assumption of Super Observer becomes non-operational since practically a measuring apparatus is a localized entity. Thus, measurement is a local event, and the synchronization problem is then manifested. The relational quantum mechanics abandons the assumption of super observer, and replaces it with the two implications.

Applying the first implication, we are able to resolve the EPR paradox~\cite{Rovelli07, Yang2018}. In that resolution, a quantum measurement should be explicitly described as observer dependent. The idea of observer-independent quantum state is abandoned since it depends on the assumption of Super Observer. By recognizing that the element of physical reality obtained from local measurement is only valid relatively to the local observer, the completeness of quantum mechanics and locality can coexist~\cite{Yang2018}. Latest experiment appears to confirm that observer-independent description of a quantum system must be
 rejected~\cite{Proietti19}. 

Applying the synchronization principle, we are able to resolve the Wigner's friend paradox and the extended version, as shown in section \ref{subsec:eWigner}. These thought experiments provide clear example for the need of information synchronization in order to achieve a consistent description of a quantum system by different observers. Ref~\cite{Brukner} shows similar idea that the assumption of observer independent fact cannot resolve the Wigner's friend type of paradox. The synchronization principle is conceptually significant since it gives the meaning of objectivity of a quantum state. The relational nature of a quantum state does not imply a quantum state is subjective. Objectivity can be defined as the ability of different observers coming to a consensus independently~\cite{Zurek03}.  

Additional conceptual implication of these results is that in quantum mechanics, information is relative. The notion of information here refers to the correlation between the observed system and another system, and is measured by the entropy of reduced density matrix of the observed system. Changes of the entropy means changes of information. A quantum process to extract information from a system must be described explicitly relative to an observer~\cite{Yang2018}. There is no absolute information to all observers in quantum mechanics, just as there is no absolute spacetime in Relativity.  

\subsection{Limitation}
There are limitations to implement the synchronization principle in certain conditions. An observer may miss the information of result of a measurement performed by another observer. Suppose a quantum system $S$ is described initially by wave function $\Psi_S(t_0)$. An observer $O_I$ performs a measurement of variable $q$ on $S$ and obtain a result $q=q_m$ at time $t_1$. Relative to $O_I$, the wave function is updated to $\Psi_S(t_1)$. However, another observer $O_E$ may not be aware of this result at time $t_2 > t_1$ due to several potential reasons. For instance, $O_I$ does not initiate the communication of measurement result, or it takes a finite period of time for the information to reach observer\footnote{Suppose it takes a period of time $\Delta t$ for the information to travel from $O_I$ to $O_E$, and $\Delta t > (t_2 - t_1)$, then at time $t_2$, $O_E$ is still not aware of the measurement result.} $O_E$, or there is error in the communication process. In any case, $O_E$ will still describe $S$ as time evolution of wave function $U_S(t_2-t_0)\Psi_S(t_0)$ where $U_S$ is the time evolution operator of $S$.

How to overcome these limitations in the communication process is an interesting problem when constructing a quantum description of a composite system with entangled subsystems. A typical procedure to construct a quantum description is to define the boundary of the composite system such that it can be approximated as an isolated system. Then, given an initial quantum state and the Hamilton operator of the system, its time evolution is described as a unitary process. If, however, an event occurs such that one of the subsystem starts to interact with another system outside the composite system and induces information exchange, a remote observer who does not know the event will describe the composite system with incomplete information, and will have inconsistent description from the oberver who knows the event. How does a local observer keep track such interaction history of the remote subsystems? This is a challenge that deserves further research.  

Note that the synchronization principle is an operational one, not a conceptual one. If the synchronization among different observers does not occur, each observer may have different descriptions of a quantum system. This is still conceptually legitimate in the context of RQM. However, if we wish to incorporate the Relativity Theory, failure of synchronization may become a conceptual issue, because having equivalent description of a physical law from different observers is a basic requirement in the Relativity Theory. How quantum measurement is described in the context of Relativity Theory? This is an interesting question to investigate given that a quantum measurement must be described as observer dependent. We speculate that the need for information synchronization in a quantum measurement is a necessary element when one wishes to combine quantum mechanics with the Relativity Theory.

\subsection{Compatibility to the Original RQM}
\label{oRQM}
As mentioned earlier, the synchronization principle is not presented in the original RQM~\cite{Rovelli96}. Instead, it is proposed in an alternative implementation~\cite{Yang2017, Yang2018} of the RQM principles. One may wonder if such principle is consistent with the original RQM. For convenience, we call the original RQM as oRQM, while the alternative implementation in Ref.~\cite{Yang2017, Yang2018} as iRQM. The following analysis shows that the synchronization principle proposed in iRQM is consistent with the ideas in oRQM. 

The synchronization process can be examined from three aspects: 1.)The observer who sends or receives measurement result; 2.)Communication of the information; 3.)The consequence of receiving the measurement result. For the first aspect, let's compare the definitions of observer in oRQM and iRQM. As mentioned in the introduction, iRQM adopts the idea from oRQM that all systems are quantum systems. The measured system $S$ and the measuring apparatus $A$ are both quantum systems. The slightly difference is that in iRQM, we distinguish the measuring apparatus $A$ and the observer $O$ as two different physical entities, and they must be locally collocated. In oRQM, they are combined as a single observer. This can be seen in section II.D of Ref~\cite{Rovelli96}, where the observer $O$ is described as both an apparatus that interacts with $S$ and also being able to ``know" the measurement outcome. This implies the observer as defined in oRQM contains something that can read, store, and compare the measurement results. Therefore an observer in oRQM is equivalent to an $A+O$ composite system in iRQM. Both definitions agree that they should be described quantum mechanically. For instance, they must follow the quantum mechanics rules of applying quantum operator to obtain measurement outcome, as described in Eq.(3) of Ref.~\cite{Rovelli96}. The key here is that there is no need to assume consciousness in the observer that influences the way quantum mechanics is interpreted. There is also no need to assume that the two observers are classical systems. The only requirement is that both entities can communicate information through physical interaction.

On the second aspect, iRQM is consistent with oRQM
on the idea that communication is achieved through physical interaction, and such physical interaction follows quantum mechanics principles, i.e., interaction outcome can be probabilistic rather than deterministic. Exactly how the two observers synchronize information is not the main focus here. However, it is achieved through physical interaction in two possible ways. 1.) Another observer $O'$ performs direct measurement on the $O+A$ composite system according to the operator $\hat{M}$ defined in Eq.(3) of Ref~\cite{Rovelli96}. 2.) Observer $O$ announces the result and observer $O'$ receives such announcement. This method was not discussed in oRQM but is an important step in the WFR experiment~\cite{FR2018}. How quantum mechanics is used to achieve the action of ``announcing" and ``receiving" a piece of information is not described. But there is no reason to assume quantum mechanics cannot describe such communication process. For instance, the information can be encoded into certain property of a photon, and the photon is emitted from observer $O$ and received by observer $O'$.

On the third aspect, upon receiving a piece of information about a measurement result, an observer must follow a quantum mechanics rule, that is, must update the quantum state according to Eq.(\ref{measureWF1}). It is crucial to note that updating the wave function upon receiving additional information rather than performing direct measurement is possible here since the wave function, or a quantum state, is just a mathematical tool to book-keep information resulting from interaction with the apparatus. There is no ontological meaning of the wave function. This is another key concept in oRQM~\cite{Rovelli18}. Once again, there is no need to assume that $O$ is a human being with any sort of consciousness, or assume it is a classical system. It is just a physical system that is programmed to follow the quantum measurement rule specified in Eq.(\ref{measureWF1}). 

In summary, although the synchronization principle is not presented in oRQM, it is not conflicting with oRQM either. It is a legitimate extension in the context of RQM.

\subsection{Conclusions}
The extended Wigner's friend thought experiment~\cite{FR2018} is analyzed in detail in the Schr\"{o}dinger picture here. The analysis in the Schr\"{o}dinger picture helps us to identify the inconsistency in the reasoning process that leads to the no-go theorem in Ref.~\cite{FR2018}. In the reasoning, an agent should make use of the available information, no more and no less. The information can be that is encoded in a known wave function, or can be obtained through direct measurement result. However, we show the reasoning process in proving the no-go theorem is inconsistent with respect to this requirement.

However, there is significant conceptual value brought up by the extended Wigner's friend thought experiment as it provides a clear example that information synchronization is needed if different observers want to have a consistent description of the same quantum system. The relational formulation of quantum measurement~\cite{Yang2018} provides two principles. First, quantum measurement needs to be described relative to the observer; Second, to ensure consistent descriptions of a quantum system, measurement outcome obtained by a local observer must be communicated to other observers. We show that these two principles can resolve the paradoxes presented in several thought experiments, including the EPR experiment, the Wigner's friend thought experiment, and its extended version.

The synchronization principle imposes a restriction to construct a quantum description of a composite system with entangled subsystems. An observer local to a subsystem needs to keep track the interaction history of other remote subsystems in order to have accurate description of the composite system or the subsystem local to the observer. This can be a challenge due to the operational limitations in the communication process. Nevertheless, it is an important problem to consider when we wish to incorporate the ideas from the Relativity Theory. We speculate that the synchronization of measurement results from different observers is a necessary component when combining quantum mechanics with the Relativity Theory.



%
%

\begin{acknowledgements}
The author sincerely thanks the anonymous reviewers of this paper for their careful reviews. The valuable comments provided help to improve the clarity of the discussions.
\end{acknowledgements}



\end{document}